\documentclass[
 amsmath,amssymb,twocolumn
 aps,prl,reprint,superscriptaddress
]{revtex4-2}

\usepackage{graphicx}
\usepackage{dcolumn}
\usepackage{bm}
\graphicspath{{Figs/}{Figsgaoerb/}} 
\usepackage[colorlinks,linkcolor=blue,anchorcolor=blue,citecolor=blue]{hyperref}
\usepackage{amsmath}
\usepackage{amssymb}
\usepackage{CJK}
\usepackage{keyval}
\usepackage{color}
\usepackage{txfonts}
\usepackage{verbatim}

\usepackage{lineno}

\begin{document}

\title{Electronic Origin of Ferromagnetic Excitations in the Candidate Spin-Triplet Superconductor CeSb$_2$
}

\author{Xiaoxiao Wang}
\affiliation{Shanghai Research Center for Quantum Sciences, Shanghai 201315, China}
\affiliation{
Laboratory of Advanced Materials, State Key Laboratory of Surface Physics,
and Department of Physics, Fudan University, Shanghai 200438, China
}%

\author{Xiaoyang Chen}%
\author{Suppanut Sangphet}%
\affiliation{
Laboratory of Advanced Materials, State Key Laboratory of Surface Physics,
and Department of Physics, Fudan University, Shanghai 200438, China
}%

\author{Yifei Fang}
\affiliation{State Key Laboratory of Ultra-intense Laser Science and Technology, Shanghai Institute of Optics and Fine Mechanics, Chinese Academy of Sciences, 201800 Shanghai, China
}

\author{Yilin Wang}
\affiliation{
School of Emerging Technology, University of Science and Technology of China, Hefei 230026, China.
}
\affiliation{New Cornerstone Science Laboratory, Hefei National Laboratory, Hefei, 230026, China}

\author{Chihao Li}
\author{Minyinan Lei}
\author{Nan Guo}
\author{Yuanhe Song}
\affiliation{
Laboratory of Advanced Materials, State Key Laboratory of Surface Physics,
and Department of Physics, Fudan University, Shanghai 200438, China
}%

\author{Rui Peng}
\email{pengrui@fudan.edu.cn}

\author{Haichao Xu}
\email{xuhaichao@fudan.edu.cn}
\affiliation{Shanghai Research Center for Quantum Sciences, Shanghai 201315, China}
\affiliation{
Laboratory of Advanced Materials, State Key Laboratory of Surface Physics,
and Department of Physics, Fudan University, Shanghai 200438, China
}%

\author{Donglai Feng}

\affiliation{New Cornerstone Science Laboratory, Hefei National Laboratory, Hefei, 230026, China}

 \begin{abstract}

The origin of quasi-one-dimensional (q1D) ferromagnetic (FM) excitations in the candidate spin-triplet superconductor CeSb$_2$ has remained unclear. Here we report an electronic mechanism for emergent q1D magnetism in the quasi-two-dimensional lattice of CeSb$_2$, revealed by angle-resolved photoemission spectroscopy (ARPES). High-resolution ARPES resolves no spin-density-wave gap on the dispersive Fermi pockets, disfavoring a nesting-driven mechanism for the q1D FM excitations. Instead, resonant ARPES reveals a pronounced selective enhancement of Ce 4$f$ spectral weight on the $C_2$-distributed Fermi pockets aligned with the Ce ladder. This observation signifies band-selective Kondo coupling that generates strongly anisotropic magnetic exchange interactions, which can naturally account for both the q1D ferromagnetic
excitations and the competing magnetic orders. Our results identify a band-selective Kondo coupling mechanism for emergent low-dimensional magnetism in correlated $f$-electron systems.

\end{abstract}

\date{\today}

\maketitle

Unconventional superconductivity is widely believed to arise from pairing mechanisms involving magnetic fluctuations rather than conventional electron-phonon coupling \cite{miyake1986spin,fujita2011progress,scalapino2012common,dean2013high}. For spin-triplet superconductivity, a phase of broad interest both for its fundamental novelty and for its potential relevance to topological quantum computing \cite{zhi2015nmr,bao2015superconductivity,balakirev2015anisotropy, yang2021spin,ran2019nearly,xu2019quasi,knafo2021low, li2019observation,ishihara2023chiral,tsutsumi2024topological}, ferromagnetic (FM) fluctuations have long been considered a critical ingredient for spin-triplet pairings. 
FM fluctuations have been observed in many candidate materials including K$_2$Cr$_3$As$_3$ and UTe$_2$ \cite{zhi2015nmr,ran2019nearly,yang2021spin,wu2015magnetism,cuono2021intrachain,sundar2019coexistence,duan2020incommensurate,knafo2021low,duan2021resonance}.
Identifying the interactions between such FM fluctuations and low-energy electronic states in candidate materials is therefore central to understanding spin-triplet pairing.

CeSb$_2$ has recently been identified as a candidate for pressure-induced spin-triplet superconductivity \cite{squire2023superconductivity,shan2025emergent}. INS measurements on CeSb$_2$ resolve well-defined quasi-one-dimensional (q1D) FM excitations that persist well above the magnetic ordering temperature, offering a possible FM channel relevant for triplet pairing \cite{shan2025emergent}. However, the microscopic origin of these q1D FM excitations is unclear. CeSb$_2$ crystallizes in a quasi-two-dimensional (q2D) lattice with nearly identical inter- and intra-ladder Ce-Ce distances [Fig.~1(a)],
so the lattice geometry cannot explain the dimensional reduction of the magnetic response. Moreover, the system does not condense into simple FM order but goes through four successive magnetic transitions below $T$ $\sim$16~K \cite{zhang2017anisotropic, zhang2022kondo, trainer2021phase}, including a commensurate AFM order with propagation vectors (-1, $\pm$$\frac{1}{6}$, 0) \cite{liu2020neutron,shan2025emergent}. Together, the q1D FM excitations in a near-isotropic q2D lattice and the multiple magnetic orders raise the possibility that the effective dimensionality of the magnetic interactions is determined not by the lattice geometry itself. This calls for a microscopic mechanism rooted in the electronic structure. 

\begin{figure}[htbp]
    \centering    \includegraphics[width=86mm]{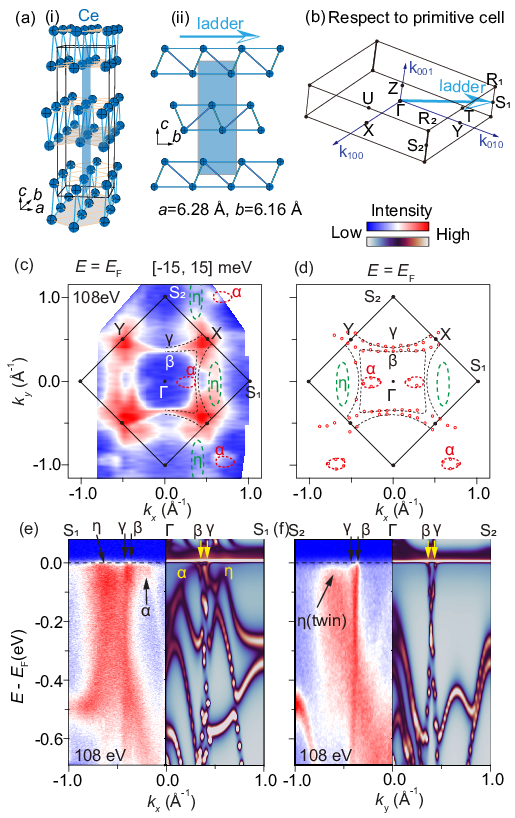}
    \vspace{-20pt}
    \caption{(a) Crystal structure of CeSb$_2$: (i) orthorhombic unit cell with the Ce sublattice highlighted; (ii) with the leg direction along the crystallographic $b$-axis (blue arrow). (b) Bulk Brillouin zone (BZ) with respect to the primitive cell. The Ce-ladder leg direction ($b$) corresponds to $\Gamma$S$_1$.
    (c) Fermi surface map measured at 6~K with 108~eV LH-polarized photons. (d) Schematic Fermi surfaces derived from (c). 
    (e) and (f) Band dispersions along the $\Gamma$S$_1$ and $\Gamma$S$_2$ directions, together with DMFT calculations.}
    \label{Fig.1}
\end{figure}

\begin{figure}[htbp]
    \centering    \includegraphics[width=86mm]{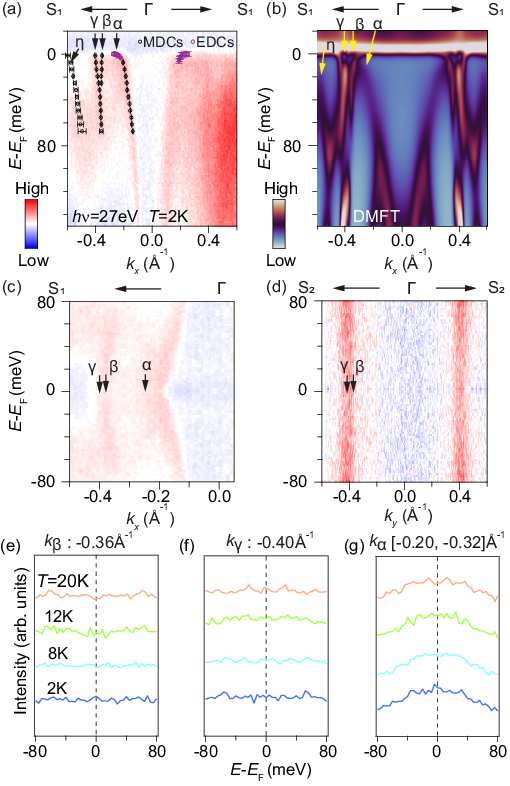}
    \caption
    {(a) Valence band dispersions near $E_{\rm{F}}$ measured at 2~K with 27~eV photons, corresponding to the $k_z$~=~$\pi$ plane. (b) Corresponding band dispersions from DMFT calculations. (c) and (d) Symmetrized energy-momentum ($E$-$k$) intensity plots along $\Gamma$S$_1$ (2~K) and $\Gamma$S$_2$ (5~K) directions, respectively, measured with 27~eV photons. (e)-(g) Temperature-dependent symmetrized energy distribution curves (EDCs) at $k_{\rm{\beta}}$, $k_{\rm{\gamma}}$ and $k_{\rm{\alpha}}$ along the $\Gamma$S$_1$ direction.}
    \label{Fig.2}
\end{figure}

Here, we address this issue by directly probing the electronic structure of CeSb$_2$ using angle-resolved photoemission spectroscopy (ARPES). We uncover coexisting near $C_4$-distributed and $C_2$-distributed Fermi surfaces. High-resolution measurements at low temperatures reveal no detectable spin-density wave (SDW) gap, which disfavors an SDW-driven scenario for the magnetic excitations. Instead, resonant ARPES suggests band-selective Kondo coupling on a subset of small $C_2$-distributed pockets aligned with the Ce ladder, generating intrinsically anisotropic magnetic exchange and competing FM/AFM tendencies. This band-selective magnetic interaction provides a natural microscopic basis for the emergence of q1D ferromagnetic fluctuations in CeSb$_2$, demonstrating how such coupling can generate q1D spin fluctuations in a q2D system.

High-quality single crystals of CeSb$_2$ were grown by the self-flux method within a graphite resistance heating system \cite{zhang2021effects}. Samples were cleaved $in$-$situ$ below 30~K along the (001) plane. ARPES measurements were performed at the synchrotron radiation facilities-Advanced Light Source (ALS), Shanghai Synchrotron Radiation Facility (SSRF), and BESSY-II under optimized experimental conditions. More details on sample growth \cite{zhang2021effects} and characterization \cite{canfield1991novel,bud1998anisotropic,zhang2017anisotropic, luccas2015charge,liu2020neutron,trainer2021phase}, dynamical mean-field theory (DMFT) calculations\cite{haule2010dynamical,haule2015free,blaha2020wien2k,haule2015exact,gull2011continuous}, and ARPES measurements were provided in Supplemental Materials (SM) Sections I-IV \cite{SM}.

\begin{figure*}[htbp]
    \centering    
    \includegraphics[width=\textwidth]{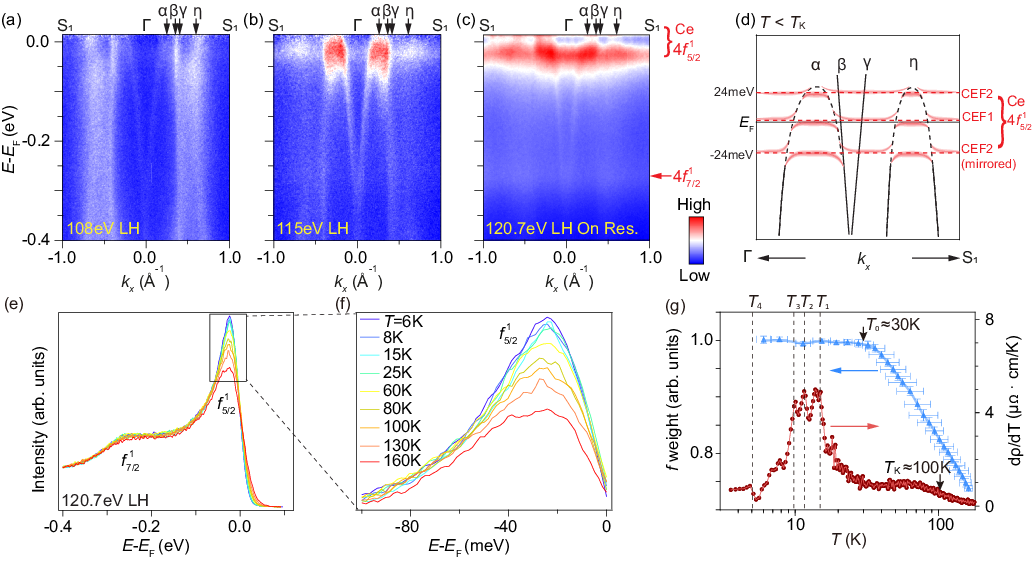}
    \caption{Photoemission intensity along $\Gamma$S$_1$ direction at 6~K with LH polarized photons at (a) 108~eV (off resonance), (b) 115~eV (pre-edge, partial resonance), and (c) 120.7~eV (on resonance, Ce $M$-edge), divided by the resolution-convoluted Fermi-Dirac distribution. 
    (d) Schematic of proposed band-selective Kondo coupling with CEF states. (e)-(f) Temperature-dependent EDCs integrated over the momentum range of panel (c). 
    (g) Spectral intensity of 4$f^1_{5/2}$ peak over the energy window [$E_{\rm{F}}$\textminus0.1~eV, $E_{\rm{F}}$+0.1~eV], normalized to its value at 6~K. The temperature derivative of resistivity d$\rho$/d$T$ is plotted to show the $T_{\rm{K}}$ and the magnetic transitions. }
    \label{Fig.3}
\end{figure*}

The electronic structure of CeSb$_2$ shows q2D character, as revealed by photon-energy dependent measurements [Fig.~S3]. The representative in-plane Fermi surfaces consist of multiple features [Figs.~1(c) and 1(d)]. An inner square-like $\beta$ pocket and an outer diamond-shaped $\gamma$ pocket are centered around the $\Gamma$ point [Figs.~1(c) and 1(d)], both preserving a nearly $C_4$ symmetry. These two Fermi surfaces cross at the X and Y high-symmetry points in a complex manner according to calculations [Fig.~S2] \cite{zhang2022kondo}; the details of these crossings are not well resolved in ARPES. In addition, two small $\alpha$ pockets and two small $\eta$ pockets are primarily located along the $\Gamma$S$_1$ direction [Figs.~1(c)-1(e)], reflecting the quasi-one-dimensional character of the Ce ladder and its intrinsic $C_2$ symmetrical distribution [Figs.~1(a)-1(b)]. Polarization-dependent measurements show that the $C_2$-distributed feature persists under both linear horizontal (LH) and linear vertical (LV) geometries [Fig.~S4], ruling out matrix-element effects.
Along both $\Gamma$S$_1$ and $\Gamma$S$_2$ [Figs.~1(e) and 1(f)], the $\beta$ and $\gamma$ bands are highly dispersive and cross $E_{\rm{F}}$ with a relatively large Fermi velocity, indicating relatively weak electronic correlation. In contrast, the $\alpha$ and $\eta$ bands show broader spectral features [Fig.~1(e)] and reduced velocities near $E_{\rm{F}}$ along $\Gamma$S$_1$ [Figs.~2(a) and 2(b)], characteristic of strong correlations, well reproduced by DMFT calculations [Figs.~1(e) and 1(f)]. 
In the ARPES spectra along $\Gamma$S$_2$ [Fig.~1(f)], the residual weight of the $\eta$ band originates from the superposition of 90$^{\circ}$-rotated structural twin domains \cite{liu2020neutron}. Micro-ARPES measurements on single-domain [Fig.~S4] confirm the absence of $\eta$ pocket along $\Gamma$S$_2$, supporting its intrinsic $C_2$-distributed character.

Given the parallel segments of the $\beta$ and $\gamma$ pockets [Fig.~1(d)], a possible Fermi surface nesting condition might give rise to SDW order and contribute to the competing magnetic orders. To test this scenario, high-resolution ARPES measurements were conducted at low temperature down to 2~K, yielding a total energy resolution of 3~meV. Symmetrized energy-momentum intensity plots along the $\Gamma$S$_1$ and $\Gamma$S$_2$ directions show no indication of gap opening at the temperature below any ordering transitions, with both $\beta$ and $\gamma$ bands dispersing to $E_{\rm{F}}$ without suppression of spectral weight. 
Upon cooling from 20~K to 2~K, the $\beta$, $\gamma$, and $\alpha$ bands disperse through $E_{F}$ along $\Gamma$S$_1$, with no detectable gap across any of the magnetic transitions [Figs.~2(e)-2(g)], within an energy resolution of 3~meV. While we cannot exclude a sub-meV gap or a partial gap, the dominance of well-defined Fermi crossings disfavors nesting as the primary driver of the complex magnetic orders below 16~K. 

Resonant ARPES measurements were performed along the $\Gamma$S$_1$ (Ce-ladder) direction by tuning the photon energy across the Ce $M$-edge. Off-resonance spectra resolve the dispersive $\alpha$, $\beta$, $\gamma$ and $\eta$ bands crossing $E_{\rm{F}}$ with no observable hybridization gap 
[Fig.~3(a)]. Upon approaching resonance [Fig.~3(b)] and at resonance [Fig.~3(c)], the spectral weight of Ce 4$f$-derived states is strongly enhanced. Distinct flat features emerge near $E_{\rm{F}}$ and at binding energy of $\sim$ 270~meV, which are attributed to the Ce 4$f^1_{5/2}$ and 4$f^1_{7/2}$ states, respectively. 
The position of the 4$f^1_{5/2}$ feature $\sim$24~meV below $E_{\rm{F}}$, differs from that in typical Ce-based heavy-fermion systems such as CeCoIn$_5$,
where the analogous feature lies $\sim$2~meV above $E_{\rm{F}}$ \cite{chen2017direct}.
It is instead reminiscent of CeSb, where the Ce 4$f$ feature appears $\sim$50~meV below $E_{\rm{F}}$, and has been interpreted as a mirrored crystal electric field (CEF) excitation \cite{jang2019direct}. 
One possible interpretation of CeSb$_2$, analogous to CeSb, is illustrated in Fig.~3(d): the primary Ce 4$f$ level (CEF1) lies at $E_{\rm{F}}$, accompanied by a CEF excitation at $\sim$24~meV above $E_{\rm{F}}$ (CEF2) and its mirrored counterpart below $E_{\rm{F}}$. (Independent confirmation of the CEF level from INS and specific heat in CeSb$_2$ is currently lacking.) In the ARPES spectra, no hybridization gap is resolved at 24~meV for any of the dispersive bands [Figs.~3(a)-3(c)], indicating that hybridization with the CEF2 state is relatively weak.
The resonant enhancement is most pronounced for the hole-like $\alpha$ and $\eta$ bands [Figs.~3(b)-3(c)], indicating band-selective Kondo hybridization/coherence and suggesting that Kondo coupling at $E_{\rm{F}}$ with the CEF1 state is finite and occurs on the Ce-ladder-aligned $C_2$-distributed Fermi pockets.  

The 4$f^1_{5/2}$ peak intensity grows monotonically upon cooling [Figs.~3(e) and 3(f)], reflecting the buildup of Kondo coherence. The flat band feature persists up to 160~K [see also Fig.~S5], indicating that heavy-quasiparticle formation develops well above the coherence temperature ($T_{K}\sim 100~K$), as commonly seen in Ce-based heavy-fermion systems \cite{chen2017direct,chen2018band,chen2018tracing,poelchen2020unexpected,wu2021revealing}. The 4$f^1_{5/2}$ weight follows a typical Kondo coherence evolution above $T_0$ $\sim$30~K but saturates below $T_0$ [Fig.~3(g)]. The sizable residual moment of $\sim$1.4~$\mu_B$/Ce at 2~K \cite{liu2020neutron}, approaching the free Ce$^{3+}$ Hund’s-rule value (2.54~$\mu_B$) and exceeding that of most Ce-based heavy fermions \cite{knopp1989magnetic,fobes2017low,raba2017determination,smidman2013neutron}, indicates incomplete Kondo screening and the persistence of substantial 4$f$ local-moment character on the Ce sites at low temperature. This relocalization-like behavior, similar to that reported in several other Ce-based heavy fermions \cite{luo2020three,aproberts2011kondo,shirer2012long,li2023photoemission,wu2023itinerant}, allows partial Kondo coherence to coexist with magnetic order.

The band-selective Kondo coupling has direct consequences for the momentum-space structure of the effective magnetic susceptibility. The resonance-enhanced Fermi surface map highlights the momentum-space distribution of Ce 4$f$ spectral weight [Fig.~4(a)]. Besides a generally enhanced intensity in the first Brillouin zone with no obvious correspondence with Fermi surfaces, the $\alpha$ and $\eta$ pockets exhibit clear band-specific enhancement as also observed in Figs.~3(b)-3(c). Since these pockets are intrinsically $C_2$-distributed, appearing along the Ce-ladder direction ($b$-axis) and absent along $\Gamma$$S_2$ [Figs.~1(c)-(f), Fig.~S4], the enhanced Kondo coupling is geometrically confined to the ladder-aligned subset of the Fermi surface. The enhanced Kondo-derived spectral weight at the Fermi wave vectors ($k_{\rm{F}}$) of the $\alpha$ and $\eta$ pockets [Figs.~4(a)-4(b)] naturally gives rise to a larger effective exchange constant between the itinerant electrons and the residual Ce 4$f$ local moments at these momenta. This band-selective anisotropy provides a direct electronic basis for highly anisotropic magnetic exchange parameters, as observed by INS \cite{shan2025emergent}. Specifically, the intra-ladder exchange $J_{\rm{\parallel}}$ dominates over the inter-ladder coupling
$J_{\rm{\perp}}$ as illustrated in Fig.~4(c), despite the quasi-2D crystal structure.

\begin{figure}[tbp]
    \centering   
    \includegraphics[width=86mm]{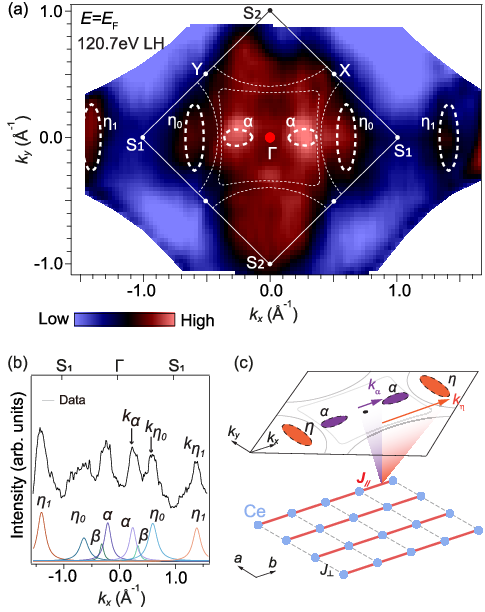}
    \caption
    {(a) Fermi surface map in the $k_x$ \textminus $k_y$ plane obtained by resonance-enhanced ARPES at 20~K with h$\nu$=120.7~eV (Ce M-edge), integrated over [$E_{\rm{F}}$ \textminus 25~meV, $E_{\rm{F}}$ + 25~meV]. 
    (b) MDC along the $\Gamma$S$_1$ direction at $E_{\rm{F}}$; The extracted Fermi wavevectors are estimated from the Lorentzian peak positions [Table II in Ref. \cite{SM}]. 
    (c) Schematic of the proposed band-selective Kondo coupling mechanism and resulting exchange anisotropy.} 
    \label{Fig.4}
\end{figure}

One plausible manifestation of the band-selective Kondo coupling could be Ruderman-Kittel-Kasuya-Yosida (RKKY)-type magnetic exchange $J(r)$ along the Ce-ladder direction $b$. Through a simplified simulation in Supplementary Materials Sec.~IX \cite{SM,litvinov1998rkky,blundell2001magnetism,rusin2017calculation}, the $\alpha$ and $\eta$ pockets favor AFM and FM nearest-neighbor exchange channels, respectively, suggesting competing magnetic interactions and the origin of q1D FM interactions. More intriguingly, within the experimental uncertainty on $k_\alpha$, the contributions from $\alpha$ and $\eta$ pockets can also produce a significant AFM channel at the third-nearest neighbor, consistent with the experimentally observed commensurate AFM order with propagation vectors (-1, $\pm$$\frac{1}{6}$, 0) \cite{liu2020neutron,shan2025emergent}.
These coexisting FM and AFM tendencies provide an electronic mechanism for both the dominant FM excitations and the complex magnetic phase diagram of CeSb$_2$ that cannot be accounted for by a single FM exchange alone \cite{shan2025emergent}.

The $f$-electron spectral weight saturates below $T_0$ $\sim$30~K, indicating incomplete Kondo screening, leaving residual Ce 4$f$ moments that remain available for magnetic ordering. 
The small size of the $\alpha$ and $\eta$ Fermi pockets suggests a low density of itinerant carriers, which may partially contribute to the incomplete screening of the Ce 4$f$ moments. In addition, CEF effects further reduce the $f$-electron spectral weight near $E_{\rm F}$. The combined effect may account for the relatively small effective mass observed in CeSb$_2$~\cite{gamble2002specific,luccas2015charge}.
Notably, the saturation of $f$-electron spectral weight occurs above the highest magnetic transition temperature $T_1$ ($\sim$~15~K) [Fig.~3(e)], suggesting the presence of short-range magnetic correlations that persist into the paramagnetic regime. This is consistent with neutron scattering results showing well-defined q1D paramagons \cite{shan2025emergent}. The coexistence of band-selective Kondo coherence and persistent spin fluctuations in the paramagnetic regime may be relevant to the emergence of pressure-induced superconductivity \cite{arndt2011spin,saxena2000superconductivity,aoki2001coexistence, ran2019nearly}.

In summary, we have identified band-selective Kondo coupling as a microscopic electronic mechanism underlying the q1D FM excitations in CeSb$_2$. The selective hybridization between Ce 4$f$ local moments and the $C_2$-distributed Fermi pockets aligned with the Ce ladder gives rise to anisotropic Kondo-mediated exchange interactions. Unlike conventional RKKY scenarios in which all conduction-electron pockets contribute uniformly to the exchange, our results reveal a band-selective mechanism in which only the ladder-aligned, Kondo-coupled pockets dominate the magnetic interactions. More broadly, our results suggest that band-selective Kondo coupling can effectively reduce the dimensionality of magnetic interactions independently of the underlying crystal lattice, providing a general electronic route to emergent q1D magnetism in correlated $f$-electron systems.

\begin{acknowledgments}
\textit{Acknowledgments --} We gratefully acknowledge the valuable discussion with Guangming Zhang, Liyang Qiu and Michael Smidman. We also thank the experimental support of Chris Jozwiak, Emile Rienks, ZhiCheng Jiang, Yichen Yang, Zhengtai Liu, Gexing Qu and Zhenhua Chen. This work is supported in part by the National Science Foundation of China under the grant Nos. 12422404, 92477206, 12274085, 92365302, the National Key R\&D Program of the MOST of China (2023YFA1406300), the New Cornerstone Science Foundation, the Quantum Science and Technology-National Science and Technology Major Project (Grant No.2021ZD0302803), and Shanghai Municipal Science and Technology Major Project (Grant No.2019SHZDZX01).
Yifei Fang acknowledges CAS Project for Young Scientists in Basic Research (YSBR-127).
The use of the Advanced Light Source, BL7.0.2, is supported by the U.S. Department of Energy, Office of Science, Office of Basic Energy Sciences under Contract No.DE-AC02-76SF00515. The low temperature and high resolution ARPES measurements were carried out at the One-cubed ARPES end-station at the BESSY II electron storage ring operated by the Helmholtz-Zentrum Berlin für Materialien und Energie. Part of this research used Beamline 03U (https://cstr.cn/31124.02.SSRF.BL03U) of the Shanghai Synchrotron Radiation Facility. Some preliminary data were taken at BL09U (https://cstr.cn/31124.02.SSRF.BL09U) of the Shanghai Synchrotron Radiation Facility.
\end{acknowledgments}

\bibliography{Ref}

@article{miyake1986spin,
  title={Spin-fluctuation-mediated even-parity pairing in heavy-fermion superconductors},
  author={Miyake, K and Schmitt-Rink, S and Varma, CM},
  journal={Physical Review B},
  volume={34},
  number={9},
  pages={6554},
  year={1986},
  publisher={APS}
}

@article{fujita2011progress,
  title={Progress in neutron scattering studies of spin excitations in high-T c cuprates},
  author={Fujita, Masaki and Hiraka, Haruhiro and Matsuda, Masaaki and Matsuura, Masato and M. Tranquada, John and Wakimoto, Shuichi and Xu, Guangyong and Yamada, Kazuyoshi},
  journal={Journal of the Physical Society of Japan},
  volume={81},
  number={1},
  pages={011007},
  year={2011},
  publisher={The Physical Society of Japan}
}

@article{scalapino2012common,
  title={A common thread: The pairing interaction for unconventional superconductors},
  author={Scalapino, Douglas J},
  journal={Reviews of Modern Physics},
  volume={84},
  number={4},
  pages={1383--1417},
  year={2012},
  publisher={APS}
}

@article{dean2013high,
  title={High-Energy Magnetic Excitations in the Cuprate Superconductor Bi 2 Sr 2 CaCu 2 O 8+ $\delta$:<? format?> Towards a Unified Description of Its Electronic and Magnetic Degrees of Freedom},
  author={Dean, MPM and James, AJA and Springell, RS and Liu, X and Monney, C and Zhou, KJ and Konik, RM and Wen, JS and Xu, ZJ and Gu, GD and others},
  journal={Physical Review Letters},
  volume={110},
  number={14},
  pages={147001},
  year={2013},
  publisher={APS}
}

@article{zhi2015nmr,
  title={NMR investigation of the quasi-one-dimensional superconductor K 2 Cr 3 As 3},
  author={Zhi, HZ and Imai, T and Ning, FL and Bao, Jin-Ke and Cao, Guang-Han},
  journal={Physical Review Letters},
  volume={114},
  number={14},
  pages={147004},
  year={2015},
  publisher={APS}
}

@article{bao2015superconductivity,
  title={Superconductivity in quasi-one-dimensional K 2 Cr 3 As 3 with significant electron correlations},
  author={Bao, Jin-Ke and Liu, Ji-Yong and Ma, Cong-Wei and Meng, Zhi-Hao and Tang, Zhang-Tu and Sun, Yun-Lei and Zhai, Hui-Fei and Jiang, Hao and Bai, Hua and Feng, Chun-Mu and others},
  journal={Physical Review X},
  volume={5},
  number={1},
  pages={011013},
  year={2015},
  publisher={APS}
}

@article{balakirev2015anisotropy,
  title={Anisotropy reversal of the upper critical field at low temperatures and spin-locked superconductivity in K 2 Cr 3 As 3},
  author={Balakirev, FF and Kong, T and Jaime, M and McDonald, RD and Mielke, CH and Gurevich, A and Canfield, PC and Bud'ko, SL},
  journal={Physical Review B},
  volume={91},
  number={22},
  pages={220505},
  year={2015},
  publisher={APS}
}

@article{yang2021spin,
  title={Spin-triplet superconductivity in K2Cr3As3},
  author={Yang, Jie and Luo, Jun and Yi, Changjiang and Shi, Youguo and Zhou, Yi and Zheng, Guo-qing},
  journal={Science Advances},
  volume={7},
  number={52},
  pages={eabl4432},
  year={2021},
  publisher={American Association for the Advancement of Science}
}

@article{ran2019nearly,
  title={Nearly ferromagnetic spin-triplet superconductivity},
  author={Ran, Sheng and Eckberg, Chris and Ding, Qing-Ping and Furukawa, Yuji and Metz, Tristin and Saha, Shanta R and Liu, I-Lin and Zic, Mark and Kim, Hyunsoo and Paglione, Johnpierre and others},
  journal={Science},
  volume={365},
  number={6454},
  pages={684--687},
  year={2019},
  publisher={American Association for the Advancement of Science}
}

@article{xu2019quasi,
  title={Quasi-two-dimensional Fermi surfaces and unitary spin-triplet pairing in the heavy fermion superconductor UTe 2},
  author={Xu, Yuanji and Sheng, Yutao and Yang, Yi-feng},
  journal={Physical Review Letters},
  volume={123},
  number={21},
  pages={217002},
  year={2019},
  publisher={APS}
}

@article{duan2020incommensurate,
  title={Incommensurate spin fluctuations in the spin-triplet superconductor candidate UTe 2},
  author={Duan, Chunruo and Sasmal, Kalyan and Maple, M Brian and Podlesnyak, Andrey and Zhu, Jian-Xin and Si, Qimiao and Dai, Pengcheng},
  journal={Physical Review Letters},
  volume={125},
  number={23},
  pages={237003},
  year={2020},
  publisher={APS}
}

@article{knafo2021low,
  title={Low-dimensional antiferromagnetic fluctuations in the heavy-fermion paramagnetic ladder compound UTe 2},
  author={Knafo, W and Knebel, G and Steffens, P and Kaneko, K and Rosuel, A and Brison, J-P and Flouquet, J and Aoki, D and Lapertot, G and Raymond, S},
  journal={Physical Review B},
  volume={104},
  number={10},
  pages={L100409},
  year={2021},
  publisher={APS}
}

@article{li2019observation,
  title={Observation of half-quantum flux in the unconventional superconductor $\beta$-Bi2Pd},
  author={Li, Yufan and Xu, Xiaoying and Lee, M-H and Chu, M-W and Chien, CL},
  journal={Science},
  volume={366},
  number={6462},
  pages={238--241},
  year={2019},
  publisher={American Association for the Advancement of Science}
}

@article{ishihara2023chiral,
  title={Chiral superconductivity in UTe2 probed by anisotropic low-energy excitations},
  author={Ishihara, Kota and Roppongi, Masaki and Kobayashi, Masayuki and Imamura, Kumpei and Mizukami, Yuta and Sakai, Hironori and Opletal, Petr and Tokiwa, Yoshifumi and Haga, Yoshinori and Hashimoto, Kenichiro and others},
  journal={Nature Communications},
  volume={14},
  number={1},
  pages={2966},
  year={2023},
  publisher={Nature Publishing Group UK London}
}

@article{tsutsumi2024topological,
  title={Topological spin texture and d-vector rotation in spin-triplet superconductors: A case of UTe 2},
  author={Tsutsumi, Yasumasa and Machida, Kazushige},
  journal={Physical Review B},
  volume={110},
  number={6},
  pages={L060507},
  year={2024},
  publisher={APS}
}

@article{sundar2019coexistence,
  title={Coexistence of ferromagnetic fluctuations and superconductivity in the actinide superconductor UTe 2},
  author={Sundar, Shyam and Gheidi, S and Akintola, K and C{\^o}t{\'e}, AM and Dunsiger, SR and Ran, S and Butch, NP and Saha, SR and Paglione, J and Sonier, JE},
  journal={Physical Review B},
  volume={100},
  number={14},
  pages={140502},
  year={2019},
  publisher={APS}
}

@article{wu2015magnetism,
  title={Magnetism in quasi-one-dimensional A2Cr3As3 (A= K, Rb) superconductors},
  author={Wu, Xian-Xin and Le, Cong-Cong and Yuan, Jing and Fan, Heng and Hu, Jiang-Ping},
  journal={Chinese Physics Letters},
  volume={32},
  number={5},
  pages={057401},
  year={2015},
  publisher={IOP Publishing}
}

@article{cuono2021intrachain,
  title={Intrachain collinear magnetism and interchain magnetic phases in Cr 3 As 3-K-based materials},
  author={Cuono, Giuseppe and Forte, Filomena and Romano, Alfonso and Ming, Xing and Luo, Jianlin and Autieri, Carmine and Noce, Canio},
  journal={Physical Review B},
  volume={103},
  number={21},
  pages={214406},
  year={2021},
  publisher={APS}
}

@article{duan2021resonance,
  title={Resonance from antiferromagnetic spin fluctuations for superconductivity in UTe2},
  author={Duan, Chunruo and Baumbach, RE and Podlesnyak, Andrey and Deng, Yuhang and Moir, Camilla and Breindel, Alexander J and Maple, M Brian and Nica, EM and Si, Qimiao and Dai, Pengcheng},
  journal={Nature},
  volume={600},
  number={7890},
  pages={636--640},
  year={2021},
  publisher={Nature Publishing Group UK London}
}

@article{squire2023superconductivity,
  title={Superconductivity beyond the Conventional Pauli Limit in High-Pressure CeSb2},
  author={Squire, Oliver P and Hodgson, Stephen A and Chen, Jiasheng and Fedoseev, Vitaly and de Podesta, Christian K and Weinberger, Theodore I and Alireza, Patricia L and Grosche, F Malte},
  journal={Physical Review Letters},
  volume={131},
  number={2},
  pages={026001},
  year={2023},
  publisher={APS}
}

@article{shan2025emergent,
  title={Emergent Ferromagnetic Ladder Excitations in Heavy Fermion Superconductor CeSb2},
  author={Shan, Zhaoyang and Jiao, Yangjie and Guo, Jiayu and Wang, Yifan and Wu, Jinyu and Zhang, Jiawen and Zhang, Yanan and Su, Dajun and Adroja, Devashibhai T and Balz, Christian and others},
  journal={Physical Review Letters},
  volume={134},
  number={11},
  pages={116704},
  year={2025},
  publisher={APS}
}

@article{zhang2017anisotropic,
  title={Anisotropic and mutable magnetization in Kondo lattice CeSb2},
  author={Zhang, Yun and Zhu, Xiegang and Hu, Bingfeng and Tan, Shiyong and Xie, Donghua and Feng, Wei and Qin, Liu and Zhang, Wen and Liu, Yu and Song, Haifeng and others},
  journal={Chinese Physics B},
  volume={26},
  number={6},
  pages={067102},
  year={2017},
  publisher={IOP Publishing}
}

@article{zhang2022kondo,
  title={Kondo entanglement in the quasi-two-dimensional heavy fermion compound CeSb 2},
  author={Zhang, Yun and Luo, Xuebing and Feng, Wei and Tan, Shiyong and Hao, Qunqing and Zhang, Qiang and Yuan, Dengpeng and Wang, Bo and Liu, Yi and Liu, Qin and others},
  journal={Physical Review B},
  volume={106},
  number={4},
  pages={045133},
  year={2022},
  publisher={APS}
}

@article{trainer2021phase,
  title={Phase diagram of Ce Sb 2 from magnetostriction and magnetization measurements: Evidence for ferrimagnetic and antiferromagnetic states},
  author={Trainer, Christopher and Abel, Caiden and Bud'ko, Sergey L and Canfield, Paul C and Wahl, Peter},
  journal={Physical Review B},
  volume={104},
  number={20},
  pages={205134},
  year={2021},
  publisher={APS}
}

@article{liu2020neutron,
  title={Neutron scattering study of commensurate magnetic ordering in single crystal CeSb 2},
  author={Liu, Benqiong and Wang, Liming and Radelytskyi, Igor and Zhang, Yun and Meven, Martin and Deng, Hao and Zhu, Fengfeng and Su, Yixi and Zhu, Xiegang and Tan, Shiyong and others},
  journal={Journal of Physics: Condensed Matter},
  volume={32},
  number={40},
  pages={405605},
  year={2020},
  publisher={IOP Publishing}
}

@article{zhang2021effects,
  title={Effects of the initial flux ratio on CeSb 2 crystal growth by a self-flux method},
  author={Zhang, Shulong and Li, Mingtao and Yang, Yilun and Zhao, Chengchun and He, Mingzhu and Hang, Yin and Fang, Yifei},
  journal={CrystEngComm},
  volume={23},
  number={29},
  pages={5045--5052},
  year={2021},
  publisher={Royal Society of Chemistry}
}

@article{jang2019direct,
  title={Direct visualization of coexisting channels of interaction in CeSb},
  author={Jang, Sooyoung and Kealhofer, Robert and John, Caolan and Doyle, Spencer and Hong, Ji-Sook and Shim, Ji Hoon and Si, Qimiao and Erten, Onur and Denlinger, Jonathan D and Analytis, James G},
  journal={Science Advances},
  volume={5},
  number={3},
  pages={eaat7158},
  year={2019},
  publisher={American Association for the Advancement of Science}
}

@article{knopp1989magnetic,
  title={Magnetic order in a Kondo lattice: A neutron scattering study of CeCu 2 Ge 2},
  author={Knopp, G and Loidl, Alois and Knorr, K and Pawlak, L and Duczmal, M and Caspary, R and Gottwick, U and Spille, H and Steglich, Frank and Murani, AP},
  journal={Zeitschrift f{\"u}r Physik B Condensed Matter},
  volume={77},
  pages={95--104},
  year={1989},
  publisher={Springer}
}

@article{fobes2017low,
  title={Low temperature magnetic structure of CeRhIn5 by neutron diffraction on absorption-optimized samples},
  author={Fobes, David M and Bauer, Eric Dietzgen and Thompson, Joe David and Sazonov, Andrew and Hutanu, Vladimir and Zhang, S and Ronning, Filip and Janoschek, Marc},
  journal={Journal of Physics: Condensed Matter},
  volume={29},
  number={17},
  pages={17LT01},
  year={2017},
  publisher={IOP Publishing}
}

@article{raba2017determination,
  title={Determination of the magnetic structure of CePt 2 In 7 by means of neutron diffraction},
  author={Raba, Matthias and Ressouche, Eric and Qureshi, N and Colin, CV and Nassif, V and Ota, S and Hirose, Y and Settai, R and Rodi{\`e}re, Pierre and Sheikin, I},
  journal={Physical Review B},
  volume={95},
  number={16},
  pages={161102},
  year={2017},
  publisher={APS}
}

@article{smidman2013neutron,
  title={Neutron scattering and muon spin relaxation measurements of the noncentrosymmetric antiferromagnet CeCoGe 3},
  author={Smidman, M and Adroja, DT and Hillier, Adrian D and Chapon, LC and Taylor, JW and Anand, VK and Singh, Ravi P and Lees, Martin R and Goremychkin, EA and Koza, MM and others},
  journal={Physical Review B},
  volume={88},
  number={13},
  pages={134416},
  year={2013},
  publisher={APS}
}

@article{chen2017direct,
  title={Direct observation of how the heavy-fermion state develops in CeCoIn 5},
  author={Chen, QY and Xu, DF and Niu, XH and Jiang, J and Peng, R and Xu, HC and Wen, CHP and Ding, ZF and Huang, K and Shu, L and others},
  journal={Physical Review B},
  volume={96},
  number={4},
  pages={045107},
  year={2017},
  publisher={APS}
}

@article{chen2018band,
  title={Band dependent interlayer f-electron hybridization in CeRhIn 5},
  author={Chen, QY and Xu, DF and Niu, XH and Peng, Rui and Xu, Hai Chao and Wen, CHP and Liu, X and Shu, L and Tan, SY and Lai, XC and others},
  journal={Physical Review Letters},
  volume={120},
  number={6},
  pages={066403},
  year={2018},
  publisher={APS}
}

@article{chen2018tracing,
  title={Tracing crystal-field splittings in the rare-earth-based intermetallic CeIrIn 5},
  author={Chen, QY and Wen, CHP and Yao, Q and Huang, K and Ding, ZF and Shu, L and Niu, XH and Zhang, Y and Lai, XC and Huang, YB and others},
  journal={Physical Review B},
  volume={97},
  number={7},
  pages={075149},
  year={2018},
  publisher={APS}
}

@article{poelchen2020unexpected,
  title={Unexpected differences between surface and bulk spectroscopic and implied Kondo properties of heavy fermion CeRh2Si2},
  author={Poelchen, Georg and Schulz, Susanne and Mende, Max and G{\"u}ttler, Monika and Generalov, Alexander and Fedorov, Alexander V and Caroca-Canales, Nubia and Geibel, Christoph and Kliemt, Kristin and Krellner, Cornelius and others},
  journal={npj Quantum Materials},
  volume={5},
  number={1},
  pages={70},
  year={2020},
  publisher={Nature Publishing Group UK London}
}

@article{wu2021revealing,
  title={Revealing the Heavy Quasiparticles in the Heavy-Fermion Superconductor CeCu 2 Si 2},
  author={Wu, Zhongzheng and Fang, Yuan and Su, Hang and Xie, Wu and Li, Peng and Wu, Yi and Huang, Yaobo and Shen, Dawei and Thiagarajan, Balasubramanian and Adell, Johan and others},
  journal={Physical Review Letters},
  volume={127},
  number={6},
  pages={067002},
  year={2021},
  publisher={APS}
}

@article{luo2020three,
  title={Three-dimensional and temperature-dependent electronic structure of the heavy-fermion compound CePt 2 In 7 studied by angle-resolved photoemission spectroscopy},
  author={Luo, Yang and Zhang, Chen and Wu, Qi-Yi and Wu, Fan-Ying and Song, Jiao-Jiao and Xia, W and Guo, Yanfeng and Rusz, J{\'a}n and Oppeneer, Peter M and Durakiewicz, Tomasz and others},
  journal={Physical Review B},
  volume={101},
  number={11},
  pages={115129},
  year={2020},
  publisher={APS}
}

@article{aproberts2011kondo,
  title={Kondo liquid emergence and relocalization in the approach to antiferromagnetic ordering in CePt 2 In 7},
  author={apRoberts-Warren, N and Dioguardi, AP and Shockley, AC and Lin, CH and Crocker, J and Klavins, P and Pines, D and Yang, Y-F and Curro, NJ},
  journal={Physical Review B},
  volume={83},
  number={6},
  pages={060408},
  year={2011},
  publisher={APS}
}

@article{shirer2012long,
  title={Long range order and two-fluid behavior in heavy electron materials},
  author={Shirer, Kent R and Shockley, Abigail C and Dioguardi, Adam P and Crocker, John and Lin, Ching H and apRoberts-Warren, Nicholas and Nisson, David M and Klavins, Peter and Cooley, Jason C and Yang, Yi-feng and others},
  journal={Proceedings of the National Academy of Sciences},
  volume={109},
  number={45},
  pages={E3067--E3073},
  year={2012},
  publisher={National Academy of Sciences}
}

@article{li2023photoemission,
  title={Photoemission signature of the competition between magnetic order and Kondo effect in CeCoGe 3},
  author={Li, Peng and Ye, Huiqing and Hu, Yong and Fang, Yuan and Xiao, Zhiguang and Wu, Zhongzheng and Shan, Zhaoyang and Singh, Ravi P and Balakrishnan, Geetha and Shen, Dawei and others},
  journal={Physical Review B},
  volume={107},
  number={20},
  pages={L201104},
  year={2023},
  publisher={APS}
}

@article{wu2023itinerant,
  title={Itinerant to relocalized transition of f electrons in the Kondo insulator CeRu4Sn6},
  author={Wu, Fan-Ying and Wu, Qi-Yi and Zhang, Chen and Luo, Yang and Liu, Xiangqi and Xu, Yuan-Feng and Lu, Dong-Hui and Hashimoto, Makoto and Liu, Hao and Zhao, Yin-Zou and others},
  journal={Frontiers of Physics},
  volume={18},
  number={5},
  pages={53304},
  year={2023},
  publisher={Springer}
}

@article{litvinov1998rkky,
  title={RKKY interaction in one-and two-dimensional electron gases},
  author={Litvinov, VI and Dugaev, VK},
  journal={Physical Review B},
  volume={58},
  number={7},
  pages={3584},
  year={1998},
  publisher={APS}
}

@book{blundell2001magnetism,
  title={Magnetism in condensed matter},
  author={Blundell, Stephen},
  year={2001},
  publisher={OUP Oxford}
}

@article{rusin2017calculation,
  title={On calculation of RKKY range function in one dimension},
  author={Rusin, Tomasz M and Zawadzki, Wlodek},
  journal={Journal of Magnetism and Magnetic Materials},
  volume={441},
  pages={387--391},
  year={2017},
  publisher={Elsevier}
}

@book{gamble2002specific,
  title={Specific heat and transport properties of the light rare-earth diantimonides},
  author={Gamble, Brian Keith},
  year={2002},
  publisher={Clemson University}
}

@article{arndt2011spin,
  title={Spin fluctuations in normal state CeCu 2 Si 2 on approaching the quantum critical point},
  author={Arndt, Julia and Stockert, Oliver and Schmalzl, Karin and Faulhaber, Enrico and Jeevan, Hirale S and Geibel, Christoph and Schmidt, Wolfgang and Loewenhaupt, Michael and Steglich, Frank},
  journal={Physical Review Letters},
  volume={106},
  number={24},
  pages={246401},
  year={2011},
  publisher={APS}
}

@article{saxena2000superconductivity,
  title={Superconductivity on the border of itinerant-electron ferromagnetism in UGe2},
  author={Saxena, SS and Agarwal, P and Ahilan, K and Grosche, FM and Haselwimmer, RKW and Steiner, MJ and Pugh, E and Walker, IR and Julian, SR and Monthoux, P and others},
  journal={Nature},
  volume={406},
  number={6796},
  pages={587--592},
  year={2000},
  publisher={Nature Publishing Group UK London}
}

@article{aoki2001coexistence,
  title={Coexistence of superconductivity and ferromagnetism in URhGe},
  author={Aoki, Dai and Huxley, Andrew and Ressouche, Eric and Braithwaite, Daniel and Flouquet, Jacques and Brison, Jean-Pascal and Lhotel, Elsa and Paulsen, Carley},
  journal={Nature},
  volume={413},
  number={6856},
  pages={613--616},
  year={2001},
  publisher={Nature Publishing Group UK London}
}

@article{canfield1991novel,
  title={Novel Ce magnetism in CeDipnictide and Di-Ce pnictide structures},
  author={Canfield, Paul C and Thompson, JD and Fisk, Z},
  journal={Journal of Applied Physics},
  volume={70},
  number={10},
  pages={5992--5994},
  year={1991},
  publisher={AIP Publishing}
}

@article{bud1998anisotropic,
  title={Anisotropic magnetic properties of light rare-earth diantimonides},
  author={Bud’ko, Sergey L and Canfield, PC and Mielke, CH and Lacerda, AH},
  journal={Physical Review B},
  volume={57},
  number={21},
  pages={13624},
  year={1998},
  publisher={APS}
}

@article{luccas2015charge,
  title={Charge density wave in layered La 1- x Ce x Sb 2},
  author={Luccas, Roberto F and Fente, A and Hanko, J and Correa-Orellana, Alexandre and Herrera, E and Climent-Pascual, Esteban and Azpeitia, J and P{\'e}rez-Casta{\~n}eda, Tom{\'a}s and Osorio, MR and Salas-Colera, E and others},
  journal={Physical Review B},
  volume={92},
  number={23},
  pages={235153},
  year={2015},
  publisher={APS}
}

@article{haule2010dynamical,
  title={Dynamical mean-field theory within the full-potential methods: Electronic structure of CeIrIn 5, CeCoIn 5, and CeRhIn 5},
  author={Haule, Kristjan and Yee, Chuck-Hou and Kim, Kyoo},
  journal={Physical Review B},
  volume={81},
  number={19},
  pages={195107},
  year={2010},
  publisher={APS}
}

@article{haule2015free,
  title={Free energy from stationary implementation of the DFT+ DMFT functional},
  author={Haule, Kristjan and Birol, Turan},
  journal={Physical Review Letters},
  volume={115},
  number={25},
  pages={256402},
  year={2015},
  publisher={APS}
}

@article{blaha2020wien2k,
  title={WIEN2k: An APW+ lo program for calculating the properties of solids},
  author={Blaha, Peter and Schwarz, Karlheinz and Tran, Fabien and Laskowski, Robert and Madsen, Georg KH and Marks, Laurence D},
  journal={The Journal of Chemical Physics},
  volume={152},
  number={7},
  pages={074101},
  year={2020},
  publisher={AIP Publishing}
}

@article{haule2015exact,
  title={Exact double counting in combining the dynamical mean field theory and the density functional theory},
  author={Haule, Kristjan},
  journal={Physical Review Letters},
  volume={115},
  number={19},
  pages={196403},
  year={2015},
  publisher={APS}
}

@article{gull2011continuous,
  title={Continuous-time Monte Carlo methods for quantum impurity models},
  author={Gull, Emanuel and Millis, Andrew J and Lichtenstein, Alexander I and Rubtsov, Alexey N and Troyer, Matthias and Werner, Philipp},
  journal={Reviews of Modern Physics},
  volume={83},
  number={2},
  pages={349--404},
  year={2011},
  publisher={APS}
}

@misc{SM,
  note = {See Supplemental Material}
}
\bibliographystyle{apsrev4-2}

\end{document}